# TWR-MCAE: A Data Augmentation Method for Through-the-Wall Radar Human Motion Recognition

Weicheng Gao, *Graduate Student Member, IEEE*, Xiaopeng Yang, *Senior Member, IEEE*, Xiaodong Qu, *Member, IEEE*, and Tian Lan, *Member, IEEE*

*Abstract*—To solve the problems of reduced accuracy and prolonging convergence time of through-the-wall radar (TWR) human motion due to wall attenuation, multipath effect, and system interference, we propose a multilink auto-encoding neural network (TWR-MCAE) data augmentation method. Specifically, the TWR-MCAE algorithm is jointly constructed by a singular value decomposition (SVD)-based data preprocessing module, an improved coordinate attention module, a compressed sensing learnable iterative shrinkage threshold reconstruction algorithm (LISTA) module, and an adaptive weight module. The data preprocessing module achieves wall clutter, human motion features, and noise subspaces separation. The improved coordinate attention module achieves clutter and noise suppression. The LISTA module achieves human motion feature enhancement. The adaptive weight module learns the weights and fuses the three subspaces. The TWR-MCAE can suppress the low-rank characteristics of wall clutter and enhance the sparsity characteristics in human motion at the same time. It can be linked before the classification step to improve the feature extraction capability without adding other prior knowledge or recollecting more data. Experiments show that the proposed algorithm gets a better peak signal-to-noise ratio (PSNR), which increases the recognition accuracy and speeds up the training process of the back-end classifiers.

## I. Introduction

ULTRAWIDEBAND (UWB) through-the-wall radar (TWR) uses the penetrating ability of low-frequency electromagnetic waves to detect human targets behind the wall and is widely used in urban combat, anti-terrorist conflicts, disaster rescue, criminal investigation, search, and rescue [1].

Manuscript received 29 July 2022; revised 20 September 2022; accepted 8 October 2022. Date of publication 10 October 2022; date of current version 24 October 2022. This work was supported in part by the National Natural Science Foundation of China under Grant 61860206012, Grant 62101042, and Grant 61901441; in part by the Natural Science Foundation of Chongqing under Grant cstc2020jcyj-msxmX0768 and Grant cstc2021jcyj-msxmX0339; and in part by the Young Teachers of Beijing Institute of Technology: Research on Indoor Target Imaging Method. *(Corresponding author: Tian Lan.)*

The authors are with the Beijing Institute of Technology Chongqing Innovation Center, Chongqing 401120, China, also with the School of Information and Electronics, Beijing Institute of Technology, Beijing 100081, China, and also with the Key Laboratory of Electronic and Information Technology in Satellite Navigation, Ministry of Education, Beijing 100053, China (e-mail: joeybg@126.com; xiaopengyang@bit.edu.cn; xdqu@bit.edu.cn; tlan@bit.edu.cn).

Digital Object Identifier 10.1109/TGRS.2022.3213748

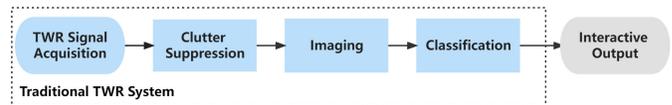

Fig. 1. Traditional TWR human motion recognition system.

In these fields, human motion recognition is one of the most challenging topics [1]. Due to the effects of walls, such as attenuation, refraction, and multipath effects [2], which cause significant distortion of the echo signal, the recognition accuracy decreases significantly, and the computational time increases.

As shown in Fig. 1, a complete TWR human motion recognition system consists of four steps: 1) signal acquisition; 2) clutter suppression; 3) imaging; and 4) classification [3]. The presence of wall clutter, systematic interference, and multipath effects in TWR data leads to poor accuracy and slow convergence. Because of this, TWR imaging needs a new data processing strategy to improve the recognition accuracy and training speed of the back-end classifiers [4].

For UWB TWR human motion recognition work, the existing methods have undergone two stages of development: 1) classical and 2) intelligent signal processing. The classical signal processing methods mainly use the fitting or decomposition algorithms to convert the image matrix into feature vectors, and then use the statistical decision methods for classification [1]. The scenario implemented by the methods at this stage is relatively simple, with poor performance, but strong interpretability [5]. Intelligent signal processing can be divided into two categories. The first category is based on sparse and low-rank modeling or traditional machine learning, including support vector machine (SVM) [6], time delay estimation method (TDOE) [7] based on the orthogonal matching tracking (OMP) algorithm, and synchronous OMP algorithm based on cross-validation (CV-CSOMP) [8]. These methods have better recognition results [9], but their generalization and inference performance are still limited when dealing with complex image data of motion state scenes [10], [11]. The second category is based on the new generation of neural network theories, including autoencoder networks (AENs) [12], convolutional neural networks (CNNs) [13], long short-term memory networks (LSTMs) [14], multilayer perceptrons (MLPs) [15], and simple probabilistic graph-based neural networks (PGMs) [16]. Due to their powerful scene generalization and feature extraction capabilities, the neural-network-based algorithms are considered the best alternatives to the original motion recognition methods. However, its training process consumes a large amount of time.

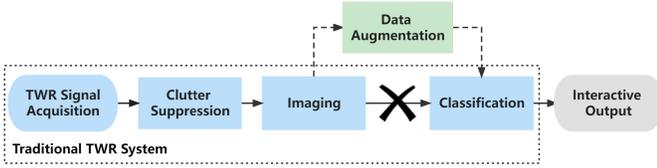

Fig. 2. New TWR human motion recognition system.

In this article, a data augmentation method between the imaging and classification steps is proposed to extract and enhance the human motion features to improve the back-end classifiers' recognition accuracy and training speed without adding a prior knowledge or reacquiring data, as shown in Fig. 2. Specifically, this article proposes a multilink convolutional auto encoding neural network (TWR-MCAE), based on the improved coordinate attention mechanism and adaptive weight learning, for through-the-wall human motion feature extraction in real scenes. The method introduces a parallel generative network architecture by combining multiscale information in 2-D image and the sparse and low-rank properties in physical context to enhance the motion features in range-time map (RTM) and Doppler-time map (DTM). Experiments demonstrate that the proposed data augmentation method achieves a higher peak signal-to-noise ratio (PSNR) and effectively improves the recognition accuracy and convergence speed of the existing classifiers.

The rest of the article is organized as follows. In Section II, we describe the signal model of TWR human motion imaging. In Section III, we introduce the structure, principle, and realization of TWR-MCAE. Section IV gives the experimental verification of the reliability, efficiency, structural optimality, and practical value of the proposed method, respectively. Finally, the conclusion is given in Section V.

## II. TWR Human Echo Model

Fig. 3 gives the geometric model of TWR human motion detection, in which the transmitting and receiving antennas of the radar are close to the wall. The signal is emitted from the transmitting antenna and refracted through the wall. Then it enters the free space. After being reflected by the target, the return wave is refracted through the free space to the wall and finally returns to the receiving antenna. The geometric model in Fig. 3 shows that the echoes in the current scenario need to introduce a wall compensation relative to the detection scenario in free space. The system uses a stepped frequency waveform as the transmit signal, which can be expressed as follows:

$$S(t) = \sum_{k=0}^{K-1} \text{rect}\left(\frac{t - T/2 - kT}{T}\right) \exp(j2\pi (f_0 + k\Delta f)t) \quad (1)$$

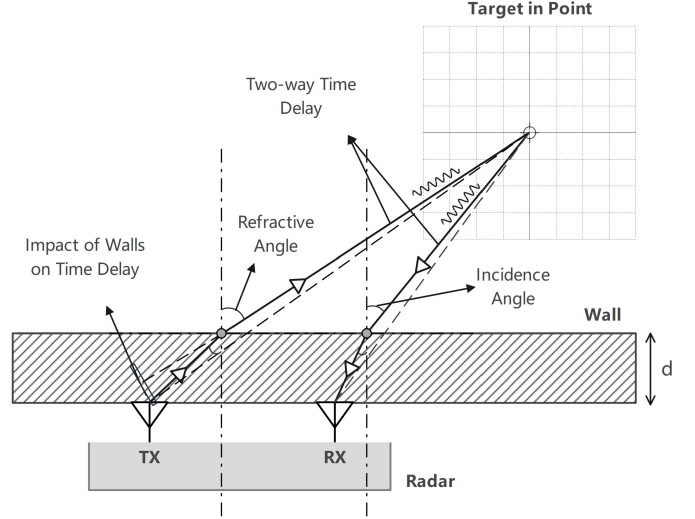

Fig. 3. Geometric model of electromagnetic propagation of TWR.

where $T$ represents the duration of each frequency point, $K$ is the total number of frequency points, $f_0$ is the starting frequency, and $\Delta f$ is the step size [17]. rect() represents the gate function and is defined as follows:

$$\text{rect}(t) = \begin{cases} 1, & t \in \left[-\frac{1}{2}, \frac{1}{2}\right] \\ 0, & t \in \left(-\infty, -\frac{1}{2}\right) \cup \left(\frac{1}{2}, \infty\right). \end{cases} \quad (2)$$

Assuming that the human target can be divided into strong scattering points, the number of the strong scattering points is $P$ [18]. Then the received echoes can be expressed as follows:

$$S_r(t) = \sum_{p=1}^{P} a_p s(t - \tau_p) + S_W(t) + S_N(t) \quad (3)$$

where $a_p$ is the echo intensity coefficient of the $p$th scattering point, $\tau_p$ denotes the echo time delay of the $p$th scattering point, and $S_W(t)$ and $S_N(t)$ denote the wall echo and noise, respectively [19].

For most TWR systems, the echo we receive is the frequency-domain data. Therefore, we directly perform local oscillation wave mixing and low-pass filtering on the echo signal to achieve coherent demodulation. The baseband radar echoes are sampled at intervals $T$. It is assumed that there are $M$ periods of echo data. After the above signal processing steps, we can reconstruct the echo data into an $M \times K$ dimensional matrix $\mathbf{S}_r$. Then one element $S_r(m, k)$ of $\mathbf{S}_r$ can be expressed as follows:

$$S_r(m, k) = \sum_{p=1}^{P} a_p(m, k) e^{-j2\pi (f_0 + k\Delta f)\tau_p} + S_W(m, k) + S_N(m, k)$$
$$m = 1, \ldots, M, k = 1, \ldots, K \quad (4)$$

where $S_W$ and $S_N$ denote the discrete reconstruction matrices of wall echoes and noise, respectively, $m$ represents the index in the slow time dimension, and $k$ represents the index in the fast time dimension. By the inverse fast

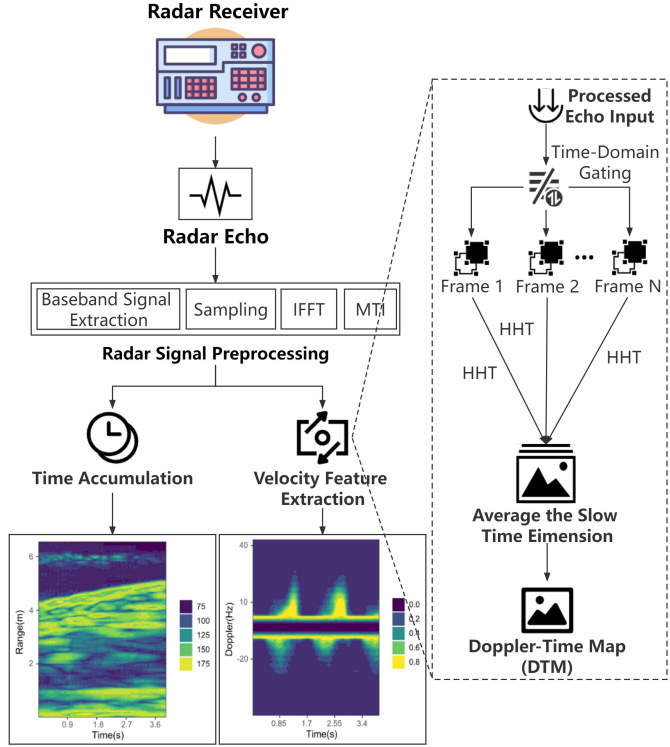

Fig. 4. Flowchart of signal preprocessing.

Fourier transform (IFFT) $\Phi_{r,\text{ori}}$ along the matrix $S_r$ fast time dimension, one of its elements $\Phi_{r,\text{ori}}(m, r)$ can be expressed as follows:

$$\Phi_{r,\text{ori}}(m, r) = \frac{1}{2\pi} \sum_{k=0}^{K-1} s(m, k) e^{j\frac{2\pi}{N}rk}$$
$$= \frac{1}{2\pi} \sum_{k \in T} s(m, k) e^{j\frac{2\pi}{N}rk} + \frac{1}{2\pi} \sum_{k \in W} s(m, k) e^{j\frac{2\pi}{N}rk}$$
$$+ \frac{1}{2\pi} \sum_{k \in N} s(m, k) e^{j\frac{2\pi}{N}rk}$$
$$r = 0, \ldots, N - 1, \quad m = 1, \ldots, M \quad (5)$$

where $r$ is the index of the number of grids being divided in the detection range. $T$, $W$, and $N$ are the subscript sets of human motion features, wall clutter, and noise subspaces, respectively. To eliminate the effect of clutter and keep back the target motion information, the moving target display (MTI) technique is used [20]. We use the difference in the spectrum between the Doppler effect of the moving target and the clutter and use blocked-band filter to suppress the clutter spectrum so as to extract the target motion information. The time-domain echo matrix after MTI processing is rewritten as follows:

$$\Phi_r = \Phi_{r,\text{ori,end}} - \Phi_{r,\text{ori,start}} \quad (6)$$

where $\Phi_{r,\text{ori,end}}$ is the submatrix consisting of the last $M - 1$ slow time periods of $\Phi_{r,\text{ori}}$, and $\Phi_{r,\text{ori,start}}$ denotes the submatrix consisting of the first $M - 1$ slow time periods of $\Phi_{r,\text{ori}}$. Define $\phi_r$ as one frame of $\Phi_r$.

In TWR human motion recognition, the existing common methods include using RTM, DTM, or combining the two for recognition [1]. As shown in Fig. 4, after MTI processing, the target echo is accumulated into a matrix in slow time. Then the echo matrix is normalized, gray-scale, and pseudo-red green blue (RGB) mapped. The final image is the RTM of human motion. On the other hand, we first perform time-domain gating on the preprocessed RTM within the range of human motion. Then Hilbert–Huang transform (HHT) is performed on every effective range unit of the time-selected RTM to obtain a time–frequency matrix corresponding to a plurality of single frequency points. HHT is defined as follows:

$$\widehat{\phi_r}(t) = \mathcal{H}\{\phi_r\} = h(t) * \phi_r(t)$$
$$= \int_{-\infty}^{\infty} \phi_r(\tau) h(t - \tau) d\tau = \frac{1}{\pi} \int_{-\infty}^{\infty} \frac{\phi_r(\tau)}{t - \tau} d\tau \quad (7)$$

where $h(t) = (1/\pi t)$ is the kernel of HHT filter response function. Then we average the time–frequency matrix of each single frequency point to obtain the Hilbert DTM. The ordinate of DTM represents the Doppler frequency offset of human motion, and the abscissa represents the slow time. Both RTM and DTM can be used for the data augmentation method to improve classifiers' recognition accuracy and convergence speed. Therefore, the numerical matrix corresponding to the image to be enhanced is all represented by $\Phi_r$.

## III. TWR-MCAE Method for Data Augmentation

As shown in Fig. 5, the proposed TWR-MCAE data augmentation method consists of four main modules and some common network structures, including the following.

1) *Data Preprocessing Module:* The data preprocessing module separates the coupled wall subspace, human motion feature subspace, and noise subspace from the TWR echo data and generates three links for neural network processing.
2) *Coordinate Attention Module:* The coordinate attention module enables the algorithm to consciously remember and understand more information about human motion and ignore image blocks containing wall clutter and noise.
3) *Learnable Iterative Shrinkage Threshold Module:* The learnable iterative shrinkage threshold (LISTA) module improves the resolution and feature contour of the block-focused RTM or DTM and effectively enhances the contrast of the regions in the image concerning human motion features.
4) *Adaptive Weighting Module:* The adaptive weighting module uses larger weights to emphasize the feature of human motion and smaller weights to make wall echoes and noise in the image less noticeable.

In addition, the data augmentation method also includes a polymerization module, a superposition module, and a validation module. These modules jointly generate the final enhanced map and guide the training of the network. The input and output structures, processing methods, hyperparameter definitions, and connection modes of every module are described in detail below.

### A. Data Preprocessing Module

In this article, the singular value decomposition (SVD) method is used to achieve the separation of wall clutter,

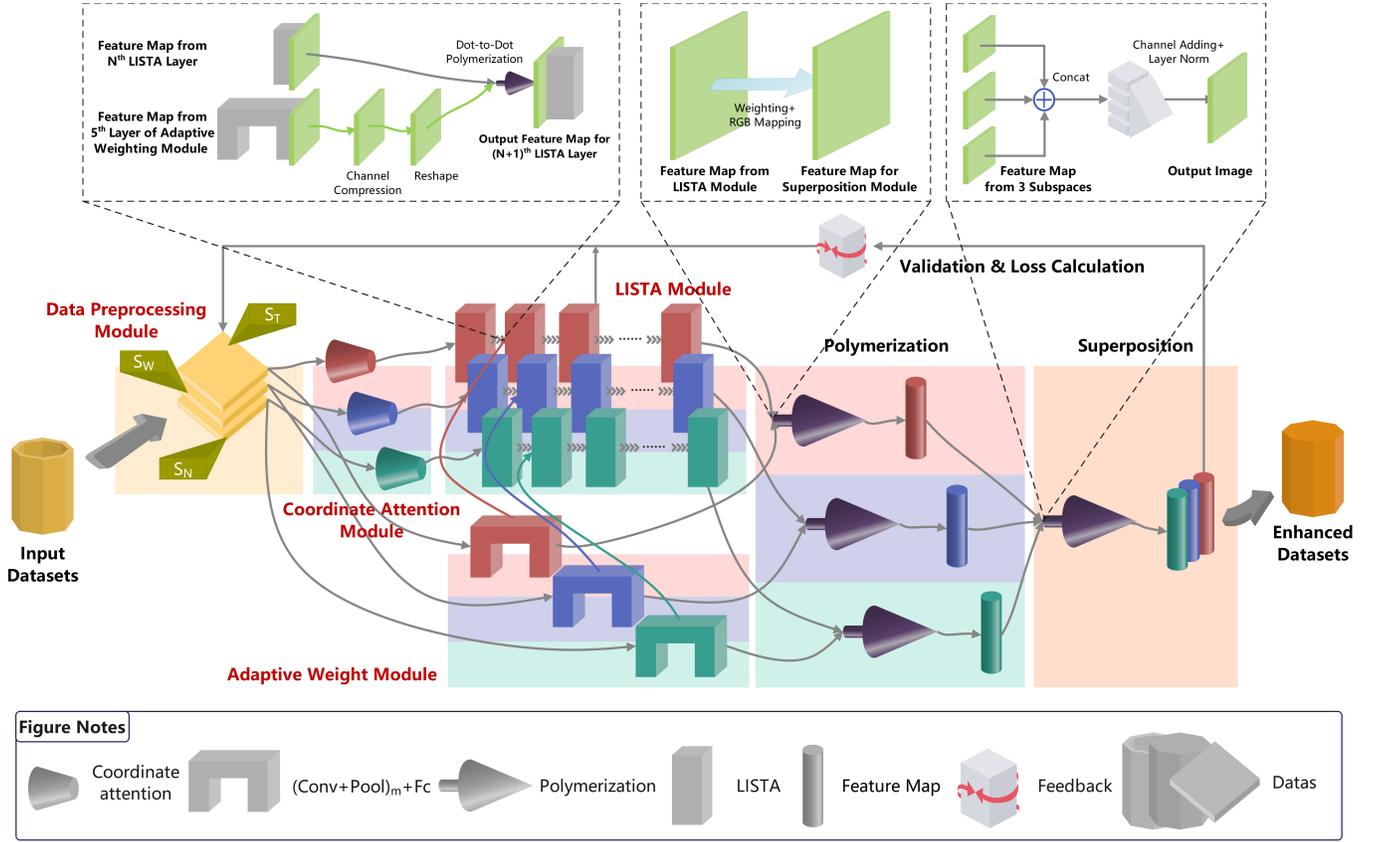

Fig. 5. Structure of the TWR-MCAE method.

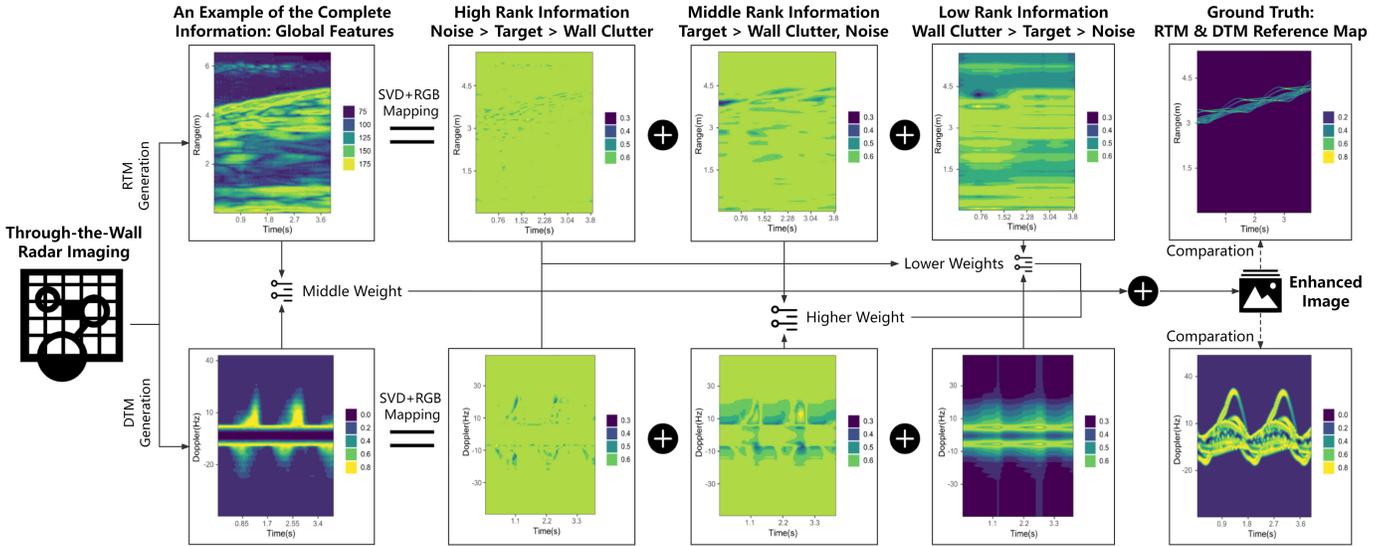

Fig. 6. Schematic of the data preprocessing module.

human motion features, and noise subspace. Its principle and purposes are shown in Fig. 6 [19]. The radar echo data are decomposed into three subspaces by the SVD method. The output images of each subspace are obtained by pseudo-RGB mapping. According to Fig. 6, the impact of noise and wall clutter can be diminished if a bigger weight is applied to the middle-rank image and a smaller weight to both the low-rank image and the high-rank image. This effectively highlights the features of human motion. Specifically, the results of SVD decomposition can be expressed as follows:

$$\boldsymbol{\Phi}_r = \boldsymbol{USV}^H = \sum_{i \in \boldsymbol{Wa}} \boldsymbol{\Phi}_{r,i} + \sum_{i \in \boldsymbol{Ta}} \boldsymbol{\Phi}_{r,i} + \sum_{i \in \boldsymbol{No}} \boldsymbol{\Phi}_{r,i}$$
$$= \sum_{i \in \boldsymbol{Wa}} \sigma_i \boldsymbol{u}_i \boldsymbol{v}_i^H + \sum_{i \in \boldsymbol{Ta}} \sigma_i \boldsymbol{u}_i \boldsymbol{v}_i^H + \sum_{i \in \boldsymbol{No}} \sigma_i \boldsymbol{u}_i \boldsymbol{v}_i^H \quad (8)$$

where $\sigma_i$ represents the singular values of the echo matrix after SVD decomposition. $\boldsymbol{u}_i$ and $\boldsymbol{v}_i$ represent the left and right chord vectors of the echo matrix after SVD decomposition,

**Algorithm 1** SVD Subspace Separation Method

**Input**: Echo matrix $\boldsymbol{\Phi}_r$, singular value $\sigma_i$, left and right vector $\boldsymbol{u}_i$, $\boldsymbol{v}_i$.
**Output**: Three subspaces $\boldsymbol{\Phi}_r^{ta}$, $\boldsymbol{\Phi}_r^{wa}$, and $\boldsymbol{\Phi}_r^{no}$.
Initializing $\boldsymbol{\Phi}_r^{ta}$, $\boldsymbol{\Phi}_r^{wa}$, and $\boldsymbol{\Phi}_r^{no} = \varnothing$
$\alpha = E(a_p)$
/* E is the mathematical expectation.           */
**for** every $\sigma_i$ **do**
  $\sigma_d = E(\sigma_i) - \alpha\sqrt{D(\sigma_i)}$
  /* Threshold of the wall subspace.            */
  **if** $\sigma_i < \sigma_d$ **then**
    $\sigma_i' = \sigma_i$
  **end**
  **else**
    $\boldsymbol{\Phi}_r^{wa} = \boldsymbol{\Phi}_r^{wa} + \sigma_i' \boldsymbol{u}_i \boldsymbol{v}_i^{\mathrm{H}}$
    $\sigma_i' = 0$
    /* Separation of the wall subspace,
       where H is the Helmut transformation.    */
  **end**
Solve:
$$\arg\min \mathrm{AIC}(i) = N\log\frac{\left(\left(\frac{1}{M-i}\right)\sum_{m=i+1}^{M}\sigma_m\right)^{M-i}}{\Pi_{m=i+1}^{M}\sigma_m} + \frac{1}{2}(2M - i)i\log N$$

  /* Threshold of the noise subspace.           */
  **if** $\sigma_i > \sigma_m$ **then**
    $\sigma_i' = \sigma_i$
  **end**
  **else**
    $\boldsymbol{\Phi}_r^{no} = \boldsymbol{\Phi}_r^{no} + \sigma_i' \boldsymbol{u}_i \boldsymbol{v}_i^{\mathrm{H}}$
    $\sigma_i' = 0$
    /* Separation of the noise subspace.        */
  **end**
  $\boldsymbol{\Phi}_r^{ta} = \boldsymbol{\Phi}_r^{ta} + \sigma_i' \boldsymbol{u}_i \boldsymbol{v}_i^{\mathrm{H}}$
  /* The remaining is the human motion feature
     subspace.                                  */
**end**

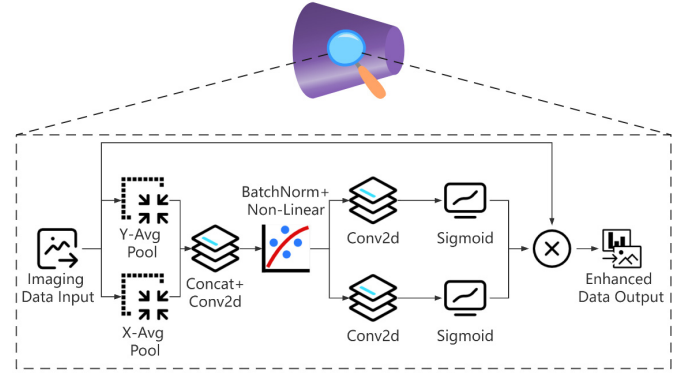

Fig. 7. Improved coordinate attention mechanism for through-wall imaging.

respectively, $U$ and $V$ are the matrices composed by $\boldsymbol{u}_i$ and $\boldsymbol{v}_i$, respectively, and $H$ denotes the Hermitian transform. The range subscripts $Ta$, $Wa$, and $No$ in the summation expression represent the target, wall clutter, and noise subspaces, respectively. The three matrix subcomponents $\boldsymbol{\Phi}_r^{ta}$, $\boldsymbol{\Phi}_r^{wa}$, and $\boldsymbol{\Phi}_r^{no}$ of the output represent the target echo, wall echo, and noise subspaces, respectively. The specific solution procedure is given by the Algorithm 1.

*B. Coordinate Attention Module*

The input of the coordinate attention module is the RTM or DTM after data preprocessing, and the output is the RTM or DTM after chunking and weighting in both slow time, range or Doppler dimensions, which aims to make the algorithm consciously remember and understand more information about human motion and ignore the image blocks containing wall clutter and noise. The structure of the attention module for through-the-wall human motion imaging is shown in Fig. 7.

After data preprocessing, a frame with 4 s of human motion data will generate three blocked RTMs or blocked DTMs that correspond to the human motion feature subspace, wall clutter subspace, and noise subspace, including $\boldsymbol{\Phi}_r^{ta}$, $\boldsymbol{\Phi}_r^{wa}$, and $\boldsymbol{\Phi}_r^{no}$, respectively. All three subspaces are compressed and mapped into $C \times H \times W$ pseudocolor maps, which are divided into three links and used as input to the three coordinate attention modules. Here, we define the $X-$direction as the range or Doppler, the $Y-$direction as the slow time axis in TWR image, and the channel and spatial scales of the image $x_c$ input to the coordinate attention mechanism are $C$ and $H \times W$, respectively. Averaging pooling is done for the $X-$direction and $Y-$direction of the image. The input image is aggregated into two separate directions of perceptual feature mapping along the vertical and horizontal directions, respectively, to obtain scales $C \times H \times 1$ and $C \times 1 \times W$, respectively, with two sets of data $z_c^h(h)$ and $z_c^w(w)$ [21] as follows:

$$z_c^h(h) = \frac{1}{W}\sum_{0\leq i<W} x_c(h,i) \tag{9}$$

$$z_c^w(w) = \frac{1}{H}\sum_{0\leq j<H} x_c(j,w). \tag{10}$$

The algorithm then two-dimensionally mixes and convolves the perceptual feature maps $z_c^h(h)$ and $z_c^w(w)$ in both the directions to obtain a set of data with scale $C/k_r \times 1 \times (W + H)$, where $k_r$ is the ratio of the number of channels. The data are then batch normalized and nonlinearized using the sigmoid activation function to compress the pixel magnitude to the $[0, 1]$ interval. The processing result is then split into two branches of $C/k_r \times H \times 1$ and $C/k_r \times 1 \times W$. A 2-D convolution with the ratio of input and output channel numbers of $1/k_r$ is applied to each of the two branches. It obtains two sets of feature mappings with embedding direction-specific information at scales of $C \times H \times 1$ and $C \times 1 \times W$, respectively, which are then encoded into two attentional feature mappings using the sigmoid activation function, each of which captures the distant dependencies of the input feature map along 1-D direction. The value of $C \times 1 \times W$ is determined by the image size of the input network, and the value of $k_r$ needs to be adjusted with the value of $C \times 1 \times W$ to obtain the best feature extraction performance. In this article, $k_r = 3$. Finally, the encoded two attention feature maps are fused with the input image of the coordinate attention module using the point-by-point multiplication to obtain an image data output with scale $C \times H \times W$ [22]. The human motion information can be effectively stored in the generated attention feature maps.

### C. LISTA Module

The input of the LISTA module is the output feature map of each link after chunked by the coordinate attention module, and the output is the enhanced data after feature extraction and image reconstruction. The purpose is to improve the resolution of the RTM or DTM. The structure of the LISTA module is shown in Fig. 8.

Specifically, let the input image of the LISTA module be $x_l$, the set of learnable parameters be $\boldsymbol{LP} = [\theta, \boldsymbol{W}_d, \boldsymbol{S}]$, the weight matrix be defined as $\boldsymbol{W}_e = (1/L)\boldsymbol{W}_d^{-1}$, the bias matrix be $\boldsymbol{S} = \boldsymbol{I} - \boldsymbol{W}_d^T \boldsymbol{W}_d$, and $\theta$ be a soft threshold. The input image is first processed by a 2-D convolution with a convolution kernel size of $3 \times 3$ and an equal number of input and output channels to obtain a series of image fragments after disassembly in spatial dimensions. The series of fragments are multiplied with the weight matrix $\boldsymbol{W}_e$, respectively, and the iterative shrinkage threshold function $F_\theta$ is defined as the activation function of the LISTA module [23] as follows:

$$F_\theta = |x - \theta|\,\text{sgn}(x) \qquad (11)$$

where $x$ represents the pixel point magnitude of the input activation function image, and sgn() is the sign function. Each pixel point magnitude of $F_\theta$ acting on the image fragment is nonlinearized, and the nonlinearized fragment matrix is multiplied with the bias matrix $\boldsymbol{S}$. Finally, the fragment matrices are added to obtain each image fragment after one LISTA enhancement process. Finally, all the fragments are stitched into a feature-enhanced image with the input image scale $C \times H \times W$ using deconvolution. The feature-enhanced image $z_l$, which is consistent with the input image scale $C \times H \times W$, is used as the output of one LISTA module as follows:

$$z_l = \text{deconv}(F_\theta(\boldsymbol{S}\,\text{conv}(x_l) + \boldsymbol{W}_e\,\text{conv}(x_l))). \qquad (12)$$

Cascading multiple layers of LISTA modules, both the input and output feature maps are scaled as $C \times H \times W$. Each layer of LISTA module processing gives a finer feature enhancement result than the previous layer of the LISTA module output. Finally, the results of each link's image enhancement are sent out and wait for the next step in the fusion process.

### D. Adaptive Weight Module

The input of the adaptive weight module is the RTM or DTM after data preprocessing, and the output is the corresponding weight size of each subspace in RTM or DTM, which aims to enable the algorithm to highlight human motion information by assigning larger weights and to reduce the intensity of wall echoes and noise in the image by assigning smaller weights. The structure of our proposed adaptive weighting module for through-the-wall imaging is shown in Fig. 9.

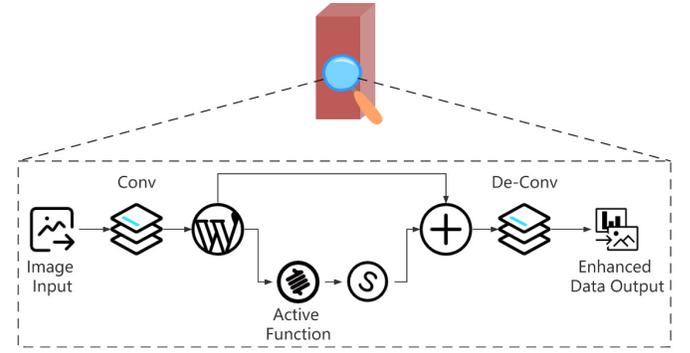

Fig. 8. Structure of LISTA module.

Specifically, similar to the preprocessing process of the coordinate attention module, all three subspaces after data preprocessing are compressed and mapped into a pseudo-RGB map of size $227 \times 227 \times 3$ [24], which are the input of the three adaptive weight modules and processed by an eight-layer deep CNN. The first layer is the convolutional layer, with the basic structure of convolution, rectified linear unit (ReLU) activation function, and pooling. The input scale of the convolution operation is $227 \times 227 \times 3$, containing 96 convolution kernels of $11 \times 11 \times 3$, padding = 0, and stride = 4. The scale size of its output feature map is $55 \times 55 \times 96$, and then it is processed by ReLU function nonlinearization, and the kernel size of the pooling operation is $3 \times 3$, padding = 0, and stride = 2. The scale size of the output feature map is $27 \times 27 \times 96$. Similarly, the second to the eighth layers are designed, and the detailed structure and parameter settings of each layer are shown in Fig. 9. The output scale size of the last layer is $1 \times 1 \times 1$, which represents the size of the weight occupied by each subspace. Let the outputs of the last fully connected layer of the adaptive weight module for the three links be $q_{(a_1)}$, $q_{(a_2)}$, and $q_{(a_3)}$, respectively, and calculate as follows:

$$z_{a_1} = \frac{q_{a_1}}{q_{a_1} + q_{a_2} + q_{a_3}} \qquad (13)$$

$$z_{a_2} = \frac{q_{a_2}}{q_{a_1} + q_{a_2} + q_{a_3}} \qquad (14)$$

$$z_{a_3} = \frac{q_{a_3}}{q_{a_1} + q_{a_2} + q_{a_3}} \qquad (15)$$

where $z_{a_1}$, $z_{a_2}$, and $z_{a_3}$ are the results of the output weights of the three adaptive weighting modules. Finally, we establish a link from the fifth layer feature map of the adaptive weight module to the output feature map of every LISTA module. Multiply and aggregate the two feature maps point by point and input them to the next LISTA module.

### E. Overall Network Structure and Loss Function

The input of the network is one RTM or DTM accumulated over a period of 4 s, and the output is the enhanced image of the input data. As shown in Fig. 5, after data preprocessing, the network is divided into a total of six branches, corresponding to two branches per subspace, one of which is processed by the coordinate attention module and the 12−layer cascaded LISTA module to obtain the enhanced images $z_{l_1}$, $z_{l_2}$, and $z_{l_3}$ for each link. The other one compresses the image scale to $227 \times 227 \times 3$ by scale transformation, and it is processed by the adaptive weight module to obtain the weight sizes $z_{a_1}$, $z_{a_2}$, and $z_{a_3}$ for each link. The six links' outputs from the three subspaces are fused using weighted summation to obtain the final data-enhanced output of the network as follows:

$$\mathbf{z}_o = \sum_{\xi=1}^{3} Z_{a_\xi} \mathbf{Z}_{l_\xi} \tag{16}$$

where $\xi$ is the scale of the subspaces. The feature enhancement effect of the network is evaluated using the perceptual function as a loss function. The mean square error (MSE) loss function can make the reconstruction result have a high PSNR, but it lacks high-frequency information and seems to have an overly smoothed texture. The perceptual loss function [25] is defined as follows:

$$L_\xi^{\text{LISTA},j}(\hat{\mathbf{z}}, \mathbf{z}) = \frac{1}{\text{CHW}} \left\| \text{LISTA}_{\xi,j}(\hat{\mathbf{z}}) - \text{LISTA}_{\xi,j}(\mathbf{z}) \right\|_2^2 \tag{17}$$

where $j$ represents the $j$th layer of the LISTA module part of the network, LISTA() represents the image after processing in the first $j$ layers of the LISTA module, $\hat{z}$ is the image we need to enhance, the source is the real data under the system settings, and $z$ is the target image, the source is the simulation dataset or real dataset with prominent features under the same scenario. $L_\xi^{\text{LISTA},j}(\hat{\mathbf{z}}, \mathbf{z})$ represents the Euclidean distance between the image to be processed and the target image [26]. Then the overall training loss function of the network is

$$\text{Loss}(\hat{\mathbf{z}}, \mathbf{z}) = \sum_{\xi=1}^{3} \sum_{j=1}^{12} z_{a_\xi} L_\xi^{\text{LISTA},j}(\hat{\mathbf{z}}, \mathbf{z}). \tag{18}$$

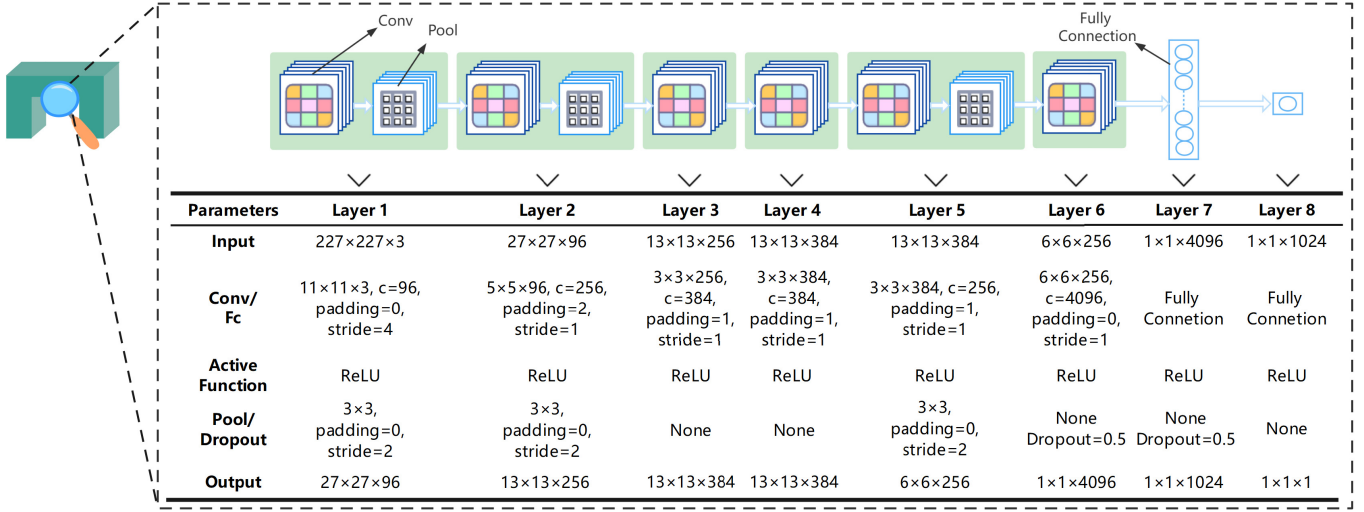

Fig. 9. Structure and detailed parameter settings of the adaptive weight module.

Both RTM and DTM contain human motion features, wall clutter, and background noise. Therefore, datasets are established for RTM and DTM, and the above methods are used for training, respectively. When it is necessary to enhance a new image with the network, first determine whether the image is an RTM before Doppler processing or a DTM after processing, and then input it into the corresponding trained network model for inference output.

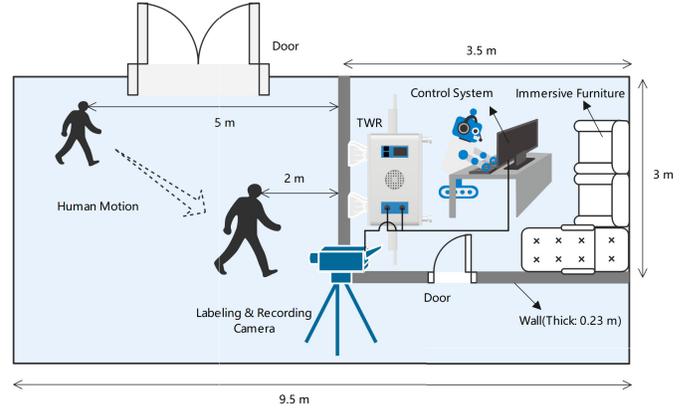

Fig. 10. Schematic of the single-station TWR.

## IV. EXPERIMENTAL VERIFICATION

In this section, we first present the designed TWR system and the human motion datasets for the experiment. Then, we test the PSNR performance of the proposed data augmentation method, compare it to the existing methods, and verify the optimality of the method's structural design. Finally, we check to see whether the data augmentation algorithm can improve the recognition accuracy and the training speed of the back-end classifiers.

### A. Experimental Setups and Dataset Construction

In this article, we use a single-transmission, single-receiver UWB TWR to collect the datasets. Fig. 10 shows the schematic of the system. For the UWB module part, we designed a 2-D scanning penetrating radar hardware and software system. To reflect the performance of the proposed method in the data augmentation task, we specially build this set of lightweight micro power TWR system to obtain poor quality data, which better simulates the challenges in practical application scenarios. Among them, the signal processing terminal of the radio frequency (RF) module is a commercial handheld vector network analyzer (VNA), Keysight N9923A [28]. The control system is connected to a UWB Vivaldi antenna fixed to the 2-D guide rail system, which is also connected to the host computer with a set of coaxial cables and forms the detection part of the whole system. The UWB TWR is placed close to the wall, and the human movement distance is in the range of 2–5 m behind the wall. Fig. 11 shows the scenario. To show a clearer and complete feature contour and simulate the real scene, we try to slow down the speed of human motion and increase the amplitude when collecting data. In the signal acquisition module, we use sliding average and interchannel averaging with a window length of 5 [29]. Finally, the images are generated using the methods mentioned in Section II, processed by pseudo-RGB mapping and contrast enhancement to obtain the data shown in the example in Fig. 12 [30]. The RTM and DTM information can be obtained in relatively complex wall and human motion environments using the current TWR systems. After data augmentation, we introduce several existing classifiers to verify their effectiveness. The detailed parameters of the radar system are given in Table I. In RTM, there are two spikes. One spike shows a relatively high modality, and the other shows a low-rank feature map. Similarly, in DTM, there are also human motion features with relatively high frequency but low energy, and low-frequency peaks with relatively high energy [27]. These two spikes approximately represent the moving body behind the wall and the wall clutter, respectively.

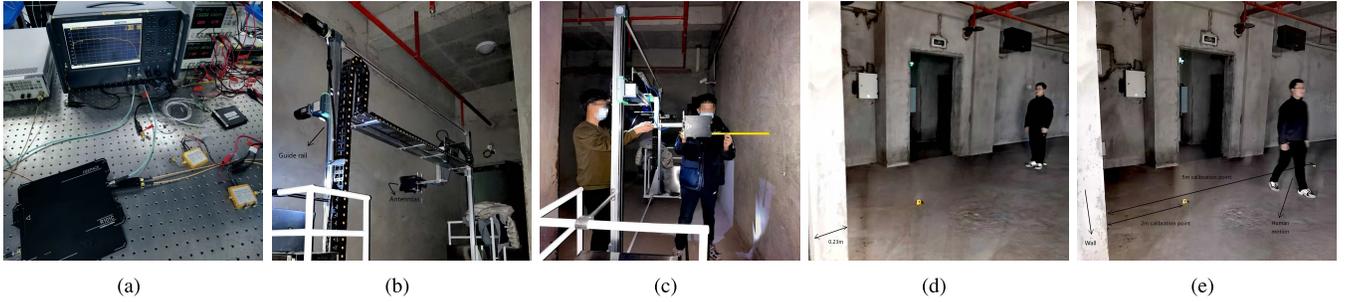

Fig. 11. Experimental scenarios. (a) Antenna testing. (b) Radar. (c) Antenna settings. (d) Measurement of human motion distance. (e) Example of the experimental process: a data acquisition of human parallel to radar walking.

### B. PSNR Comparison Experiments of Images Before and After Data Augmentation

Assuming that the through-the-wall human motion image is a $m \times n$ matrix, and $\text{MAX}_I$ is the point that represents the maximum value of the focus intensity (point spread function) $I$ in image. Combined with the design concept of the loss function, the PSNR value is used to evaluate the reconstruction quality of the proposed data augmentation method. The definition of PSNR is as follows:

$$\begin{aligned}\text{PSNR} &= 10 \cdot \log_{10}\left(\frac{\text{MAX}_I^2}{\text{MSE}}\right) \\ &= 20 \cdot \log_{10}\left(\frac{\text{MAX}_I}{\sqrt{\text{MSE}}}\right)\end{aligned} \quad (19)$$

TABLE I
PARAMETERS OF THE RADAR SYSTEM

| Parameters | Value |
|---|---|
| Antenna Transceiver Spacing | (SISO) 0.15 m |
| Antenna Operation Control | Two-Dimensional Guide Rail (Static) |
| Center Frequency | 1.5 GHz (0.5 − 2.5 GHz) |
| Frequency Step | 10 MHz |
| Sampling Points | 1024 |
| Sampling Period | 4 s |
| Wall Thickness | 0.23 m |
| Wall Material | Double Layer Concrete Brick Wall |
| Estimation of Wall $\epsilon_r$ | 7.4 |
| Human Range of Motion | $2 \sim 5$ m |
| Human Motion State | 7 (One for Empty Space) |
| Antenna Height | 1.5 m |

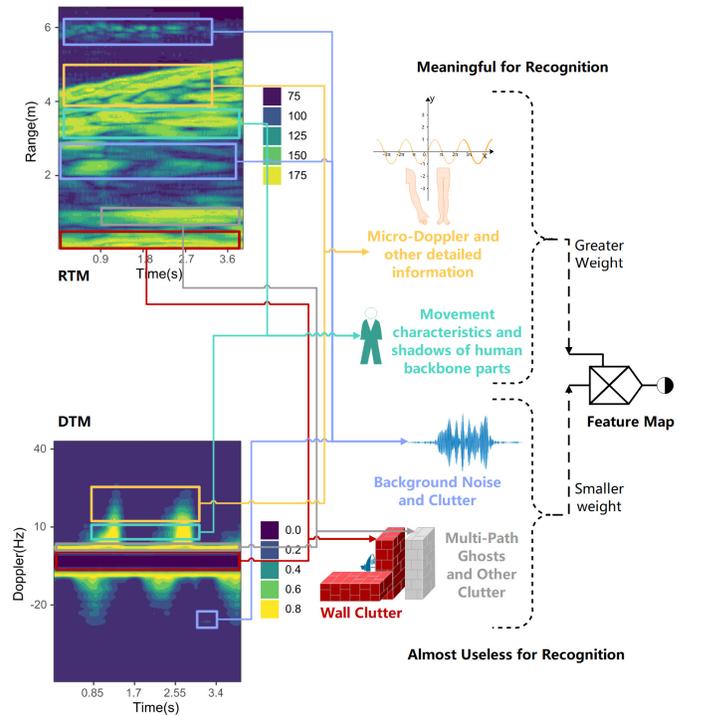

Fig. 12. Examples of RTM, DTM, and the interpretation about how the data augmentation method work on them.

where

$$\text{MSE} = \frac{1}{mn}\sum_{i=0}^{m-1}\sum_{j=0}^{n-1}\|\boldsymbol{I}(i,j) - \boldsymbol{K}(i,j)\|^2 \quad (20)$$

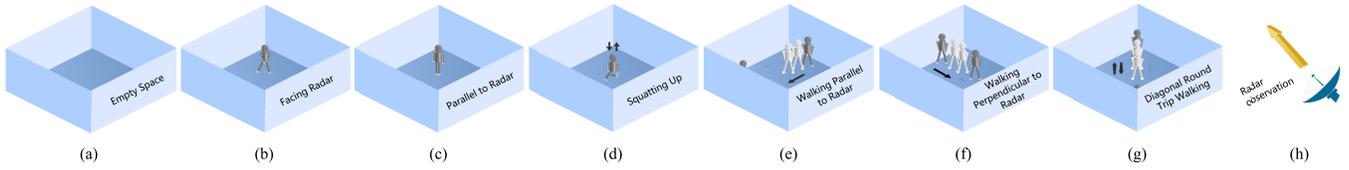

Fig. 13. Human motion states. (a)–(g) Seven substates, one for empty space, three for human activity without changing position, and three for human activity with changing position. (h) Angle of the radar observation.

in which $K$ represents the image before data augmentation. $I$ is the image after data augmentation. The accepted evaluation standard is that the PSNR higher than 40 dB indicates that the image quality is excellent (which means very close to the original one). PNSR of 30–40 dB usually indicates that the image quality is good (which means the distortion is perceptible but acceptable). PSNR of 20–30 dB indicates poor image quality. Finally, images with PSNR lower than 20 dB are unacceptable. The following evaluation work focus on seven different states shown in Fig. 13.

Fig. 12 shows how the algorithm works on a set of real experiment data. In one RTM image, the regions near the radar and away from the human body contain wall clutter, multipath ghosting, and some noise. These regions are not meaningful for the recognition task, but their information can interfere with the classifier's decision. Therefore, a lower weight should be imposed on the clutter and noise regions. The human motion feature part of the region contains the motion features of the main body parts and some tightly coupled micro-Doppler features, whose motion status is the key information for recognition and should be forced to apply higher weights. Similarly in one DTM image, the signal corresponding to the frequency around 0 Hz is meaningless, while the relatively high-frequency information of human motion is meaningful, which can also be weighted and aggregated. Finally, the output feature map information is more helpful for the learning and convergence of the classifiers.

As shown in Fig. 14(a), when there is no human target behind the wall, the image information is mainly focused on the wall clutter. Since wall clutter is low-rank information, it reflects similar features in the unoccupied image data. Thus, with the training of the coordinate attention mechanism module in the TWR-MCAE algorithm, the wall clutter region is gradually understood by the network and assigned a lower weight. When the input parameters of the neural network are data containing an unmanned environment and six other scenes containing human information, after a large number of iterations, the neural network will automatically update the adaptive weights of the coordinate regions where the wall clutter is located. The final weights are reduced to a level where the current region can be ignored. Then, we use the multiplier to combine the weight information from the adaptive weight module and the coordinate attention LISTA module's output on image chunking. The aggregated image information matrix is the result of data augmentation. In Fig. 14(a), both the RTM and DTM information of the wall region is better suppressed after the LISTA module and adaptive weight processing, and on the contrary, the information of the space behind the wall further accentuated. Since there is no human target in the current scene, only a few noisy parts in the image information are not picked up by the neural network results and weighed down to make them less noticeable. Therefore, the method can effectively make the empty scene data cleaner.

Fig. 14(b)–(d) are three different states of through-the-wall human motion image, in which the position of the human body relative to the radar will almost not change. These include marking time facing radar, standing still parallel to the radar, and squatting up. The human body's closest point to the radar antenna is fixed at around 3 m. At this point, the radar image information mostly consists of the echo information of the stationary or slightly shaking human body with the significant wall clutter. The micro Doppler frequency and amplitude corresponding to these motion states are different, which can be distinguished by the coordinate attention and LISTA modules. The data augmentation method, like the previous unmanned space, efficiently suppresses the clutter on the walls and the majority of the noise, while also giving a clear knowledge of the coordinate region of data where the human body is located. Since the position does not change, the motion characteristics of the human body also show a certain degree of low-rank characteristics. This makes the feature extraction process of the classifiers challenging. As a result, the data augmentation method filters out some of the low-rank components in the data from human imaging but keeps the components that are most important for classification and recognition. Therefore, for these three cases of RTM and DTM, the proposed data augmentation method can still play a good role.

Similarly, Fig. 14(e)–(g) represent the image information and algorithm processing results when the human body walks parallel, perpendicular, and diagonally folded back to the wall, respectively. The RTM and DTM of walking human contain more complicated micro-Doppler characteristics and have fewer components that are misclassified during data-enhanced inference, making them easier for the algorithm to extract. In particular, the radar image information still includes near-Gaussian modeling noise, substantial wall clutter, complex distance, and frequency information of human motion in the range of 2–5 m. The suggested algorithm achieves the purpose of highlighting the human motion and suppressing the wall clutter and noise by learning the information within the human body range and gradually increasing its weight during the training phase. The experimental results show that the information in the regions with high-modal micro-Doppler features is well kept. This is because the moving human features, especially the human imaging features moving parallel to the radar antenna, have low-rank regions that look like wall clutter, which the neural network may get rid of by adding lower weights. Therefore, the method has good data augmentation performance on these three cases.

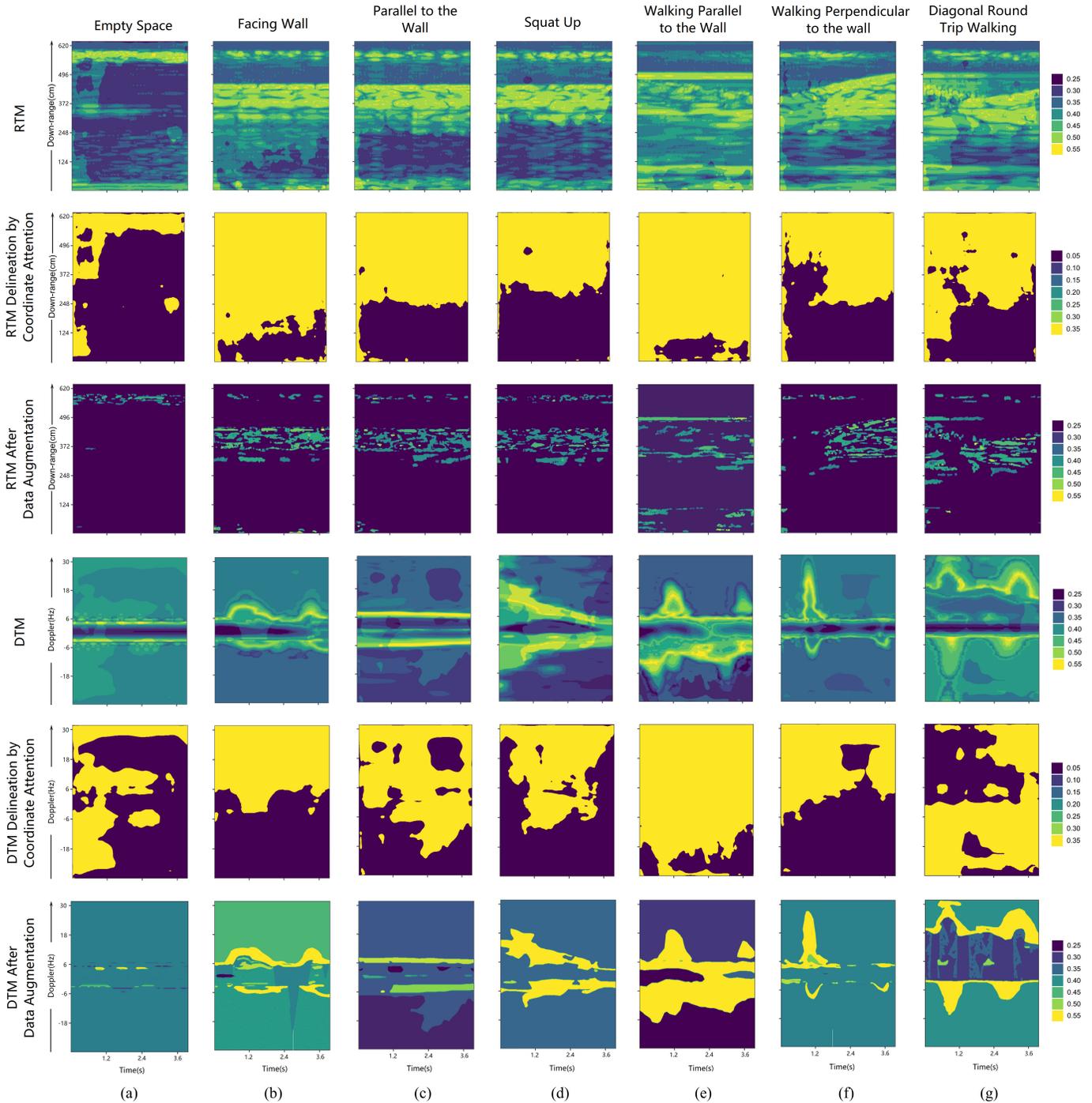

Fig. 14. RTM, DTM, and their delineation of coordinate attention regions and images after data augmentation. (a) Empty space. (b) Facing wall. (c) Parallel to the wall. (d) Squat up. (e) Walking parallel to the wall. (f) Walking perpendicular to the wall. (g) Diagonal round trip walking.

For each of the seven different states, we conduct experiments and find the PSNR before and after data augmentation, and then compare them with some previous data augmentation methods in other fields shown in Table II. Among them, linear interpolation, bicubic interpolation, and nearest neighbor (NN) interpolation are the classical image enhancement methods. L1- and L2-sparse coding are the image reconstruction methods driven by the compressed sensing theory. Rapid and accurate super image resolution (RAISR), range–$\tau$ conversion, range–max conversion, and range–Doppler (r-D) conversion are the image quality restoration algorithms referenced from other fields. Generative adversarial network (GAN), pixel-by-pixel CNN (Pixel-CNN), efficient subpixel CNN (ESPCNN), parallax attention for stereo image super-resolution network (PASSRNet), and cascade U-net are the cutting-edge methods that have been used for TWR image enhancement. The comparison shows that previous methods achieve a PSNR of 31.75–40.01 dB, and the proposed method achieves a PSNR of 40.39 dB on the real-world RTM data and 40.05 dB on the real-world DTM data. This proves that there is no existing method that can enhance RTM and DTM to a perfect level except cascade U-net. Cascade U-net can enhance RTM to a perfect level, but its ability to enhance DTM is still limited. The proposed method solves this problem and outperforms the existing methods in both RTM and DTM. It also reduces the number of network parameters compared with Cascade U-Net. As a result, the proposed TWR-MCAE algorithm has a better performance.

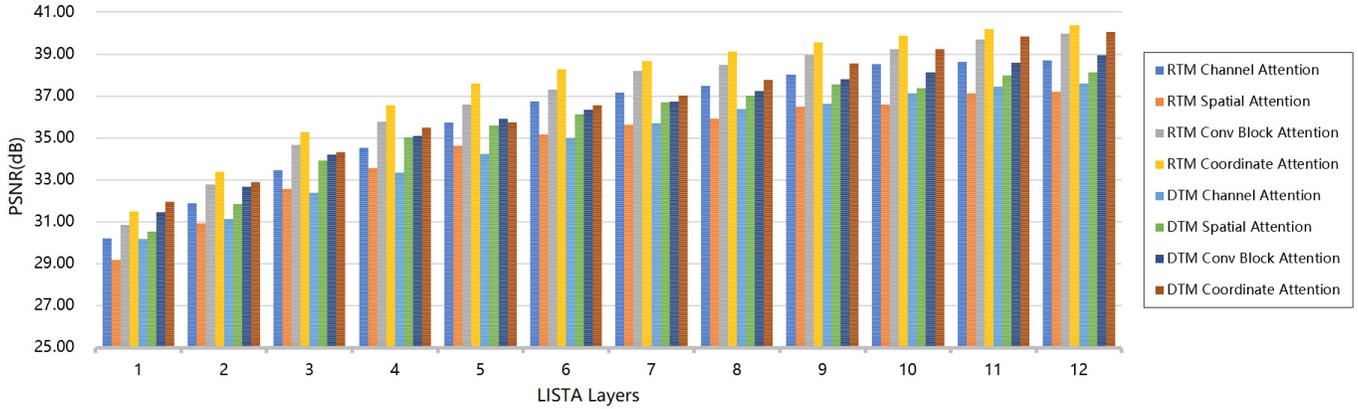

Fig. 15. HOG of PSNR comparison corresponding to different parallel layers of the LISTA module and the replacement of the attention mechanism.

TABLE II
PSNR OF TWR-MCAE AND PREVIOUS ALGORITHMS

| Name of the algorithm | RTM PSNR (dB) | DTM PSNR (dB) |
| --- | --- | --- |
| Linear Interpolation [31] | 32.12 | 31.75 |
| Bicubic Interpolation [32] | 33.35 | 31.90 |
| NN Interpolation [33] | 33.92 | 32.26 |
| L1-Sparse Coding [34] | 34.28 | 33.77 |
| L2-Sparse Coding [35] | 35.16 | 34.41 |
| RAISR [36] | 36.50 | 36.08 |
| Range-$\tau$ Conversion [8] | 35.56 | 37.13 |
| Range-Max Conversion [18] | 37.23 | 37.08 |
| R-D Conversion[1] [37] | 36.99 | 35.91 |
| GAN [38] | 37.75 | 36.80 |
| Pixel CNN [39] | 37.83 | 37.21 |
| ESPCN [40] | 38.09 | 37.43 |
| PASSRNet [41] | 39.47 | 39.06 |
| Cascade U-Net [42] | 40.01 | 39.88 |
| **TWR − MCAE** | **40.39** | **40.05** |

[1] Gaussian kernel function is used for r-D conversion.

The PSNR values of the algorithm are compared in Table III and Fig. 15 after replacing the coordinate attention module with the existing channel attention, spatial attention, or convolutional attention module. It is not difficult to find that the PSNR value of the proposed data augmentation method shows an increasing trend as the number of parallel layers of the LISTA module increases. Combined with the physical model of TWR, the PSNR of the proposed data augmentation method decreases when the number of parallel layers of the LISTA modules is less than 12. In fact, when the layers of the LISTA modules is greater than 12, the PSNR will still increase, but the increase speed is slow. At the same time, with the increase in LISTA layers, the network parameters and forward reasoning calculation will also increase. Therefore, balancing the computational cost and enhancement performance, we believe that the minimum number of LISTA layers 12, which can make both RTM and DTM achieve perfect image quality, is the best design. In addition, the PSNR of the method also decreases when the coordinate attention module is replaced with the other three commonly used attention modules. As a result, whether based on RTM or DTM training, the current method is constructed in a relatively optimal way.

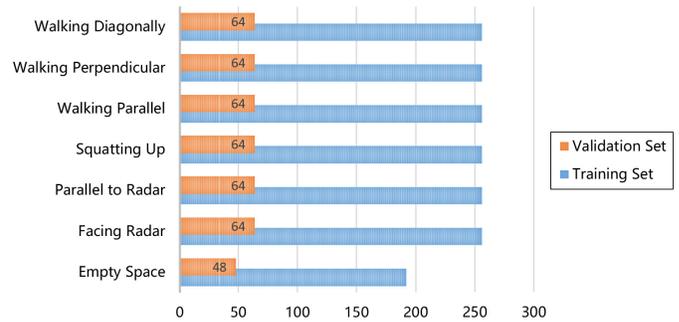

Fig. 16. Information on the number of training and validation datasets.

### C. Classification Comparison Experiments With and Without Data Augmentation

In this section, we use some traditional classifier algorithms to verify the proposed data augmentation method. The number of data in the training and validation sets is given in Fig. 16. Table IV shows the classification algorithms of the neural network that maintain the unified parameter setting during training. The comparison includes several popular classifiers that have been used in TWR human motion recognition, such as SVM, kernel function estimators with directional gradient histogram (HOG) information, classifiers based on empirical modal decomposition (EMD), and Bayesian decision, random forest (RF) classifiers, several common probabilistic graphical models (PGMs) and CNNs for classification tasks, and other algorithms that have been used by academics for TWR human motion recognition. Specifically, we verified the changes in the accuracy and convergence speed of recognition using only RTM, only DTM, and fusion these two maps with and without data augmentation. There are three methods for fusion recognition these two kinds of image information. The method based on the data layer fusion mechanism concatenates RTM and DTM by line [46]. After normalization and scaling, it is input into the classifier for recognition. The method based on the feature layer fusion mechanism inputs RTM and DTM into feature pyramid network (FPN) [47], respectively. After multi-scale feature aggregation, the two feature maps are spliced and superimposed according to the channel dimension. Then it is normalized, scaled, and input into the classifier for recognition. The method based on the decision layer fusion mechanism inputs RTM and DTM into the classifier, respectively, and the output result is the probability vector of the seven states. The probability vectors are fused by the method of maximum entropy (ME) [48], and the decision results are given. The whole comparison process is shown in Fig. 17, and the results are shown in Tables V–VII. The definition of recognition accuracy is as follows:

$$\text{Acc} = \frac{\text{TP} + \text{TN}}{\text{TP} + \text{FP} + \text{FN} + \text{TN}} \quad (21)$$

where FN: the amount of data decided as negative samples, but in fact positive samples; FP: the amount of data decided as positive samples, but in fact negative samples; TN: the amount of data decided as negative samples, but in fact also negative samples; TP: the amount of data decided as positive samples, but in fact also positive samples.

TABLE III
PSNR COMPARISON CORRESPONDING TO DIFFERENT PARALLEL LAYERS OF THE LISTA MODULE AND THE REPLACEMENT OF THE ATTENTION MECHANISM

| | LISTA Layers[1] | 1 | 2 | 3 | 4 | 5 | 6 | 7 | 8 | 9 | 10 | 11 | 12 |
|---|---|---|---|---|---|---|---|---|---|---|---|---|---|
| RTM | Channel Attention [43] | 30.22 | 31.88 | 33.46 | 34.51 | 35.74 | 36.75 | 37.16 | 37.49 | 38.02 | 38.50 | 38.63 | 38.71 |
| | Spatial Attention [44] | 29.16 | 30.91 | 32.57 | 33.57 | 34.63 | 35.18 | 35.62 | 35.92 | 36.49 | 36.61 | 37.13 | 37.21 |
| | Conv Block Attention [45] | 30.85 | 32.79 | 34.67 | 35.78 | 36.60 | 37.32 | 38.19 | 38.48 | 38.96 | 39.24 | 39.71 | 39.97 |
| | Coordinate Attention | 31.50 | 33.39 | 35.27 | 36.57 | 37.59 | 38.27 | 38.65 | 39.13 | 39.56 | 39.88 | 40.20 | 40.39 |
| DTM | Channel Attention | 30.17 | 31.14 | 32.38 | 33.35 | 34.24 | 34.99 | 35.7 | 36.39 | 36.62 | 37.13 | 37.45 | 37.61 |
| | Spatial Attention | 30.52 | 31.85 | 33.92 | 35.03 | 35.58 | 36.14 | 36.69 | 37.01 | 37.55 | 37.38 | 37.97 | 38.14 |
| | Conv Block Attention | 31.47 | 32.68 | 34.21 | 35.08 | 35.91 | 36.33 | 36.73 | 37.24 | 37.79 | 38.11 | 38.6 | 38.95 |
| | Coordinate Attention | 31.95 | 32.89 | 34.31 | 35.48 | 35.73 | 36.57 | 37.01 | 37.76 | 38.57 | 39.25 | 39.84 | 40.05 |

[1] The corresponding value in the table is PSNR, unit: dB.

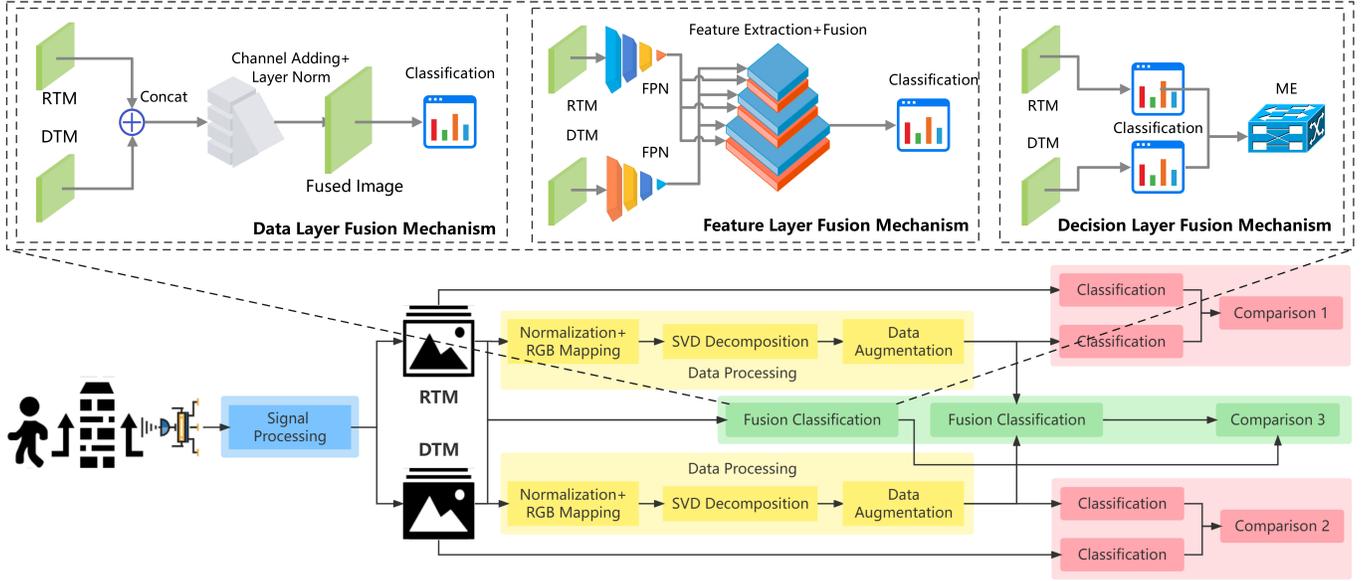

Fig. 17. Schematic of the validation process.

TABLE IV
UNIFORM PARAMETERS OF CLASSIFICATION ALGORITHMS

| Parameters | Value |
|---|---|
| Batch Size | 32 |
| Batches in Every Epoch | 54 |
| Total Epochs | 120 |
| Total Batches | 6480 |
| Initial Learning Rate | 0.005 |
| Validation Frequency(Batches) | 6 |
| Validation Times(Batches) | 1080 |
| Shuffle Frequency(Epochs) | 1 |
| Optimizer | Nesterov Momentum SGD ($\alpha = 0.9$) |
| Batch Normalization Statistics | Population |
| Whether to Train Layer by Layer | No |
| Hardware Settings | NVIDIA Tesla V100 16G |

When only RTM is used to train the classifiers, the proposed data augmentation method can effectively improve the accuracy. Among them, for the decision method of statistical signal processing, because of its relatively poor feature extraction ability, the accuracy rate with data augmentation is much higher. Some methods can exceed 8.5%. For some non-network traditional machine learning methods, the accuracy can be effectively improved by more than 4.6%. For some deep learning methods based on CNNs, the accuracy is improved by more than 5%. When using the original RTM for classification, the accuracy of all the comparison methods is less than 90%. In the RTM classification with data augmentation, both ResNet-50 and ResNet-101 methods exceed the validation accuracy of 90%. It is proven that the proposed data augmentation method can improve the recognition accuracy of RTM to a higher level. For the neural network methods, we also compared the training convergence speed with and without data augmentation. The number of convergence rounds of all the classifiers decreased. Among them, the training rounds of GoogleNet series and Resnet-101 on the original RTM exceeded 100, while the convergence speed with data augmentation increased by more than 10%. This proves the potential value of our proposed data augmentation method in system deployment. Similarly, when only DTM is used for training, the proposed data augmentation method can also effectively improve the validation accuracy and convergence speed of various classifiers. Due to the high similarity of DTM features of some motion states, the accuracy of some classifiers is less than 60%, which makes their reliability low. However, with data augmentation, the recognition accuracy of all the classifiers exceeds 65%. This proves that the proposed data augmentation method further improves the interpretability of the model. In addition, we compare the recognition accuracy and the convergence speed for the three ensemble methods based on data layer, feature layer, and decision layer fusion. Among them, the data layer fusion has the worst interpretability and the feature representation is not clear, so the recognition accuracy is relatively low among the three fusion methods. The proposed data augmentation method can still improve the accuracy and convergence speed. The feature level fusion has the strongest interpretability and clearest feature representation, so the recognition accuracy is relatively high among the three fusion methods. Since data augmentation mainly works on human motion features, the effect is the most significant in feature layer fusion classification. Taking ResNet-101, which has the highest recognition accuracy, as an example, the recognition accuracy improves by more than 3%, and the training convergence speed increases by 27%. Finally, using the data augmentation method, the GoogleNet-V4, VGG-19, and ResNet series based on decision layer fusion can all achieve a high recognition accuracy of more than 95%, and the training speed can be improved by more than 16%. This proves that the proposed data augmentation method still has a good performance improvement on the ensemble recognition method of multidimensional features.

TABLE V
RTM RECOGNITION ACCURACY OF DIFFERENT COMMONLY USED CLASSIFIERS WITH AND WITHOUT ADDING TWR-MCAE

| Name of Classifiers[1] | Validation Accuracy WODA[2] | Validation Accuracy WDA[2] | C-Epoches[3] WODA | C-Epoches[3] WDA |
|---|---|---|---|---|
| **Classification Methods in Statistical Signal Processing** | | | | |
| SVD + Statistical Decision [49] | 65.97% | 73.61% | ╲ | ╲ |
| ISVD + Statistical Decision [50] | 68.06% | 76.62% | ╲ | ╲ |
| EMD + Statistical Decision [51] | 71.30% | 79.40% | ╲ | ╲ |
| HOG + Statistical Decision [52] | 72.69% | 73.38% | ╲ | ╲ |
| **Classification Methods in Classical Machine Learning** | | | | |
| KNN [53] | 80.79% | 85.42% | ╲ | ╲ |
| PCA + SVM [54] | 82.41% | 84.95% | ╲ | ╲ |
| RPCA + RBF Kernel SVM [11] | 83.56% | 86.34% | ╲ | ╲ |
| RF [55] | 85.19% | 87.73% | ╲ | ╲ |
| **Classification Methods in Neural Network** | | | | |
| BNN [56] | 83.10% | 87.04% | 30 | 27 |
| HMM Based Network [57] | 82.87% | 85.65% | 61 | 49 |
| CRF Based Network [58] | 84.03% | 89.12% | 38 | 32 |
| GoogleNet Inception V1 [59] | 85.42% | 88.43% | 147 | 116 |
| GoogleNet Inception V2 [59] | 86.57% | 89.58% | 141 | 125 |
| GoogleNet Inception V3 [59] | 86.11% | 86.81% | 133 | 109 |
| GoogleNet Inception V4 [59] | 87.96% | 90.05% | 118 | 91 |
| VGG-16 [60] | 84.49% | 87.96% | 72 | 63 |
| VGG-19 [59] | 84.95% | 88.19% | 87 | 69 |
| ResNet-18 [13] | 87.50% | 89.81% | 53 | 45 |
| ResNet-50 [13] | 88.66% | 91.90% | 86 | 71 |
| **ResNet − 101 [13]** | **89.58%** | **93.06%** | **101** | **84** |

[1] This table is the accuracy and convergence speed results only trained and verified by RTM.
[2] 'WODA' is an abbreviation for 'Without Data Augmentation', and 'WDA' is an abbreviation for 'With Data Augmentation'.
[3] 'C-Epoches' is an abbreviation for 'Convergence Epoches'.

TABLE VI
DTM RECOGNITION ACCURACY OF DIFFERENT COMMONLY USED CLASSIFIERS WITH AND WITHOUT ADDING TWR-MCAE

| Name of Classifiers[1] | Validation Accuracy WODA[2] | Validation Accuracy WDA[2] | C-Epoches[3] WODA | C-Epoches[3] WDA |
|---|---|---|---|---|
| **Classification Methods in Statistical Signal Processing** | | | | |
| SVD + Statistical Decision | 58.56% | 65.28% | ╲ | ╲ |
| ISVD + Statistical Decision | 60.19% | 67.13% | ╲ | ╲ |
| EMD + Statistical Decision | 60.88% | 63.43% | ╲ | ╲ |
| HOG + Statistical Decision | 63.66% | 68.98% | ╲ | ╲ |
| **Classification Methods in Classical Machine Learning** | | | | |
| KNN | 59.95% | 69.21% | ╲ | ╲ |
| PCA+SVM | 67.59% | 75.93% | ╲ | ╲ |
| RPCA+RBF Kernel SVM | 71.99% | 76.85% | ╲ | ╲ |
| RF | 75.00% | 81.02% | ╲ | ╲ |
| **Classification Methods in Neural Network** | | | | |
| BNN | 78.94% | 82.64% | 36 | 31 |
| HMM Based Network | 79.86% | 84.72% | 41 | 40 |
| CRF Based Network | 82.18% | 85.19% | 35 | 28 |
| GoogleNet Inception V1 | 81.25% | 84.03% | 130 | 95 |
| GoogleNet Inception V2 | 80.56% | 86.34% | 142 | 123 |
| GoogleNet Inception V3 | 85.65% | 86.11% | 138 | 121 |
| GoogleNet Inception V4 | 85.88% | 88.89% | 126 | 104 |
| VGG-16 | 79.63% | 84.95% | 66 | 63 |
| VGG-19 | 80.32% | 87.04% | 75 | 67 |
| ResNet-18 | 83.80% | 86.81% | 71 | 59 |
| ResNet-50 | 84.72% | 85.88% | 90 | 69 |
| **ResNet − 101** | **87.27%** | **89.81%** | **110** | **92** |

[1] This table is the accuracy and convergence speed results only trained and verified by DTM.
[2] 'WODA' is an abbreviation for 'Without Data Augmentation', and 'WDA' is an abbreviation for 'With Data Augmentation'.
[3] 'C-Epoches' is an abbreviation for 'Convergence Epoches'.

TABLE VII
FUSION RECOGNITION ACCURACY OF DIFFERENT COMMONLY USED CLASSIFIERS WITH AND WITHOUT ADDING TWR-MCAE

| Name of Classifiers[1] | Validation Accuracy WODA[2] | Validation Accuracy WDA[2] | C-Epoches[3] WODA | C-Epoches[3] WDA |
|---|---|---|---|---|
| **Data Layer Fusion Mechanism** | | | | |
| EMD + Statistical Decision | 72.22% | 80.09% | ╲ | ╲ |
| HOG + Statistical Decision | 78.70% | 84.26% | ╲ | ╲ |
| PCA+SVM | 84.03% | 85.42% | ╲ | ╲ |
| RPCA+RBF Kernel SVM | 85.65% | 88.66% | ╲ | ╲ |
| RF | 87.04% | 90.51% | ╲ | ╲ |
| BNN | 89.12% | 89.58% | 44 | 34 |
| HMM Based Network | 90.28% | 93.75% | 52 | 47 |
| CRF Based Network | 91.90% | 94.21% | 46 | 33 |
| GoogleNet Inception V1 | 90.97% | 92.36% | 128 | 107 |
| GoogleNet Inception V2 | 91.67% | 93.98% | 131 | 120 |
| GoogleNet Inception V3 | 93.06% | 94.68% | 116 | 102 |
| GoogleNet Inception V4 | 93.06% | 95.14% | 112 | 94 |
| VGG-16 | 91.20% | 94.44% | 78 | 61 |
| VGG-19 | 92.59% | 93.98% | 85 | 69 |
| ResNet-18 | 92.13% | 95.14% | 66 | 64 |
| ResNet-50 | 93.75% | 95.37% | 95 | 81 |
| **ResNet − 101** | **94.44%** | **96.06%** | **107** | **88** |
| **Feature Layer Fusion Mechanism** | | | | |
| EMD + Statistical Decision | 81.94% | 87.50% | ╲ | ╲ |
| HOG + Statistical Decision | 83.10% | 86.81% | ╲ | ╲ |
| PCA+SVM | 86.81% | 90.28% | ╲ | ╲ |
| RPCA+RBF Kernel SVM | 87.96% | 91.90% | ╲ | ╲ |
| RF | 92.36% | 95.14% | ╲ | ╲ |
| BNN | 90.05% | 92.13% | 25 | 21 |
| HMM Based Network | 91.20% | 94.44% | 44 | 32 |
| CRF Based Network | 92.59% | 94.21% | 29 | 26 |
| GoogleNet Inception V1 | 87.96% | 92.36% | 101 | 79 |
| GoogleNet Inception V2 | 92.36% | 94.68% | 97 | 65 |
| GoogleNet Inception V3 | 93.29% | 94.91% | 109 | 90 |
| GoogleNet Inception V4 | 94.68% | 95.83% | 81 | 74 |
| VGG-16 | 90.51% | 92.36% | 57 | 51 |
| VGG-19 | 92.59% | 94.21% | 69 | 55 |
| ResNet-18 | 93.52% | 96.06% | 60 | 44 |
| ResNet-50 | 95.14% | 97.22% | 71 | 59 |
| **ResNet − 101** | **95.60%** | **98.61%** | **93** | **68** |
| **Decision Layer Fusion Mechanism[4]** | | | | |
| EMD + Statistical Decision | 77.55% | 79.17% | ╲ | ╲ |
| HOG + Statistical Decision | 79.86% | 82.64% | ╲ | ╲ |
| PCA+SVM | 85.42% | 87.73% | ╲ | ╲ |
| RPCA+RBF Kernel SVM | 86.11% | 89.12% | ╲ | ╲ |
| RF | 89.35% | 89.81% | ╲ | ╲ |
| BNN | 89.12% | 91.20% | 28 | 23 |
| HMM Based Network | 90.51% | 93.75% | 37 | 29 |
| CRF Based Network | 91.90% | 93.98% | 35 | 30 |
| GoogleNet Inception V1 | 90.74% | 92.82% | 105 | 84 |
| GoogleNet Inception V2 | 90.97% | 92.59% | 122 | 121 |
| GoogleNet Inception V3 | 92.13% | 95.37% | 99 | 89 |
| GoogleNet Inception V4 | 93.29% | 96.06% | 86 | 73 |
| VGG-16 | 94.21% | 95.14% | 54 | 46 |
| VGG-19 | 94.68% | 96.76% | 63 | 49 |
| ResNet-18 | 93.52% | 96.30% | 52 | 38 |
| ResNet-50 | 94.44% | 96.53% | 71 | 63 |
| **ResNet − 101** | **94.91%** | **96.99%** | **80** | **67** |

[1] This table is the accuracy and convergence speed results using 3 different fusion mechanisms.
[2] 'WODA' is an abbreviation for 'Without Data Augmentation', and 'WDA' is an abbreviation for 'With Data Augmentation'.
[3] 'C-Epoches' is an abbreviation for 'Convergence Epoches'.
[4] The model before decision-making layer fusion is consistent with the model trained separately, so the number of each model's training convergence rounds is the same, while the ensemble training is relatively faster. The accuracy is the result after fusion.

Summarizing all the experimental results, we can get the following conclusion.

1) *Reliability:* The proposed data augmentation method can effectively extract information from RTM and DTM, enhance human motion features, and reduce the interference of wall clutter and noise.
2) *Efficiency:* Compared with the existing data augmentation methods for TWR, the proposed method has the highest PSNR for human motion image.
3) *Optimality of Structural Design:* The proposed data augmentation method has higher PSNR results by replacing modules or changing network layers.
4) *High Practical Value:* The proposed data augmentation method can improve the accuracy and convergence speed of RTM only, DTM only, and three kinds of ensemble recognition methods.

## V. Conclusion

In this article, we propose a multilink auto-encoding neural network (TWR-MCAE) data augmentation method to solve the problems of decreasing accuracy and prolonging the training time of TWR human motion recognition due to wall attenuation and multipath effect. Specifically, the TWR-MCAE method is jointly constructed by an SVD-based data preprocessing module, an improved coordinate attention module, a LISTA module, and an adaptive weight module. The data preprocessing module achieves wall clutter, human motion features, and noise subspace separation. The improved coordinate attention module achieves clutter and noise suppression. The LISTA module achieves human motion feature enhancement. The adaptive weight module learns the weights and fuses the three subspaces. The TWR-MCAE can simultaneously improve the sparsity features of human motion and decrease the low-rank characteristics of wall clutter. To increase the feature extraction capability without adding more prior knowledge or gathering more data, it can be linked before the classification stage. Experiments show that the PSNR of the proposed algorithm is better than previous results. The recognition accuracy and training speed of the existing classification algorithms are both improved by the proposed method.

## Acknowledgment

For a selection of permitted open source materials related to this research, please visit the author's official website at https://weichenggaoresearch.company.site/.

The authors would like to thank Junbo Gong from the Chongqing Innovation Center, Beijing Institute of Technology, and Jiancheng Liao and Jingbo Wang from the Special Radar Laboratory (formerly: the New System Radar Laboratory), Beijing Institute of Technology, for their help in related experiments. In addition, the authors also would like to thank Zhengliang Zhu and his team for their outstanding work on "Open-Source Dataset of Human Motion Status Using IR-UWB Through-Wall Radar" and Dr. Pengyun Chen and his team for their series of outstanding work on "Ultra-Wideband Radar Human Motion Intelligent Recognition," which are the inspiration and encouragement of this work.

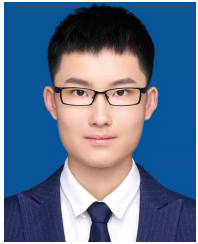

**Weicheng Gao** (Graduate Student Member, IEEE) received the B.E. degree in Xu Teli class of excellence from the Beijing Institute of Technology (BIT), Beijing, China, in 2022, where he is currently pursuing the Ph.D. degree with the Research Laboratory of Radar Technology.

He is currently an Advisor of the Radar Club, BIT, and the Peer Tutor of the Physics Foundation Class. His research interests include new system radar signal processing and interpretable machine learning, especially in through-the-wall radar human motion and gait recognition techniques.

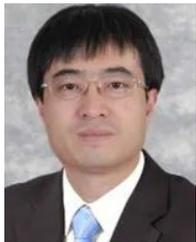

**Xiaopeng Yang** (Senior Member, IEEE) received the B.E. and M.E. degrees from Xidian University, Xi'an, China, in 1999 and 2002, respectively, and the Ph.D. degree from Tohoku University, Sendai, Japan, in 2007.

He is currently a Professor and the Ph.D. Mentor, a Secretary of the Radar Research Laboratory, Beijing Institute of Technology, Beijing, China, and the Associate Director of innovation and intelligence base of national-level discipline of new institutional radar. He mainly performs scientific research and teaching related to radar system and radar signal processing in the new institution and hosts several longitudinal topics such as the Key Points of National Natural Foundation and the National 863 Program, he had published over 200 academic papers and has 14 national invention patents, and he does special report more than ten times in major academic conferences at home and abroad.

Dr. Yang is a member of the China Electronic Society. He is the paid Director of the China Radar Industry Association, a Board Member of the China Society for Occupational and Technical Education, the Associate Director of radar chapter of the China Electronic Society, a Board Member of Young Scientist Club, a Signal Processing Chapter Member, a Sole Asian Board Member of the IEEE Aerospace Electronics System Society and the IEEE RSP, and the editorial board member of multiple domestic and foreign academic journals. He was successively awarded the Outstanding Scientific and Technical Worker of the China Electronic Society, the Outstanding Journal Dissertation Award in the electronic field of China, the IEEE and IET International Conference and the National Radar Academic Annual Meeting Excellent Dissertation Award, the Provincial and Ministerial Excellent Ph.D., the master's thesis, and the Undergraduate Bi Setting Mentorship Teacher Award. He is the Chair of the IET International Radar Conference 2020, the IEEE International Conference on Signal Information Data Processing 2019, and the Advanced Radar Signal Processing International Forum 2019 Process Committee.

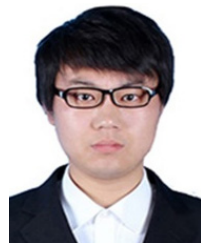

**Xiaodong Qu** (Member, IEEE) received the B.S. degree from Xidian University, Xi'an, China, in 2012, and the Ph.D. degree from the University of Chinese Academy of Sciences, Beijing, China, in 2017.

His research interests mainly include array signal processing, through-the-wall radar imaging.

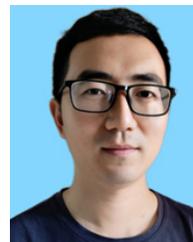

**Tian Lan** (Member, IEEE) received the B.S. degree in electromagnetic fields and wireless technology and the M.S. degree in electromagnetic fields and microwave technology from the University of Electronic Science and Technology of China, Chengdu, China, in 2011 and 2014, respectively, and the Ph.D. degree in electromagnetic fields and microwave technology from Xiamen University, Xiamen, China, in 2020.

From 2019 to 2020, he was a Senior Antenna Engineer with OPPO Company Ltd., Dongguan, China. Since 2020, he has been a Post-Doctoral Research Associate with the Beijing Institute of Technology, Beijing, China. His research areas mainly include electromagnetic inversion in ground-penetrating radar and through-the-wall radar.